\documentclass[a4paper,11pt]{article}
\usepackage{jinstpub} 
\usepackage{lineno}
\nolinenumbers
\title{\boldmath Target Development Using the Method of High-Intensity Vibrational Powder Plating (HIVIPP) at the Center for Accelerator Target Science (CATS) at Argonne National Laboratory (ANL)}
\author {C. Mohs\textsuperscript{1},}
\author{C. M{\"u}ller-Gatermann\textsuperscript{1},}
\author{M. Gott\textsuperscript{2},}
\author{J. Nolen\textsuperscript{1},}
\author{R. Gampa\textsuperscript{1},}
\author{J. Greene\textsuperscript{1}}
\affiliation{\textsuperscript{1} Argonne National Laboratory, 9600 S. Cass Ave., Lemont, IL 60439, USA \newline \textsuperscript{2}Enrichment Science and Engineering Division, 
2 Oak Ridge National Laboratory, 1 Bethel Valley Road, Oak Ridge, TN 37830 USA}

\emailAdd{jgreene@anl.gov}
\abstract{One of the primary goals of the Center for Accelerator Target Science (CATS) is to provide targets and foils in support of the ATLAS User Facility and the Low-Energy community at large. While a wide array of target production techniques are available at CATS, new methods that must be explored invariably arise. One such technique, the High-Intensity Vibrational Powder Plating (HIVIPP), was first reported in 1997 by Isao Sugai. It was developed to produce targets and stripper foils that were difficult to make by standard methods. At Argonne National Laboratory (ANL), we have successfully constructed and tested a simple system for this purpose. We have produced targets of carbon and titanium on various metal backings using the HIVIPP method. We are currently in the exciting phase of exploring the production of other elements, including isotopically enriched and radioactive material. This work is in progress and will be further detailed with specific examples.}

\arxivnumber{1234.56789} 
\begin{document}
\maketitle
\flushbottom
\section{Introduction and Motivation}

High-Intensity Vibrational Powder Plating (HIVIPP), was first reported in 1997 by Isao Sugai \cite{sugai_new}; it was developed to produce targets and stripper foils that were difficult to make by standard methods due to their high melting point and resistance to rolling. It has proven successful despite not being widely used. The idea behind this technique is to use particles' vibrational movements in a static electric field to plate target materials onto metal substrates. The aggressive movement of the particles contained between these metal backings eventually causes the particles to deposit onto the foils. This method is temperature-independent, can be performed in a vacuum or a controlled gaseous atmosphere, and only requires that the target material is conductive. 
\newline While Sugai initially employed this technique for carbon stripper foils and other difficult metals \cite{sugai_nucl}, one of our motivations for pursuing HIVIPP was to prepare isotopic carbon targets. While Carbon targets can be readily made in the CATS Facility using an E-gun evaporator source, it is a very inefficient method and not ideal for Carbon-14 due to its radioactive nature. In this regard, the HIVIPP method was an attractive alternative due to the compactness of the cell containing the radioactive powder and the high efficiency of the technique (100 percent recovery of not deposited material possible). Therefore, the plating cell and holder assembly were designed to isolate the radioactive powder in order to minimize contamination risk. In addition to Carbon, other difficult target preparations (for example, Arsenic, which can be hazardous) and the refractory metals (which cannot be readily evaporated) are also being explored as the flexibility of this method allows for a wide variety of substrates and conductive metal powder.

\section{Experimental Technique}
The principle behind the HIVIPP method is the phenomenon of vibrational motion of metallic particles via contact-charge scattering in a static electric field \cite{Tanabe}. The electric field is produced between metal plates separated by a short distance using an insulating cylindrical ring containing the particles. The kinetic motion produced in the particles by the voltage held across the plates eventually results in deposition onto the electrodes. For the present setup, the HIVIPP apparatus consists of a stack of two substrate foils held apart a short distance by a non-conductive quartz cylinder. These substrate foils can be identical or different and are connected at the top and bottom to a high-voltage power supply and ground, respectively. A small quantity of the target material to be deposited is placed inside the quartz cylinder on the bottom substrate. This powdered target material must be conductive to accept charge from the electrode at contact and experience the electric field established between the two foils with the application of high voltage across them. No attempt was made to determine the powder consumption rate for these initial studies. The substrates are held in place in recessed Macor supports to reduce static discharge and insulate the apparatus (see Fig. 1). The simple design employs removable quartz cylinders of 16 mm diameter and a height of 15 mm. One may also use larger cylinders if larger diameter targets are preferred. It was incorporated from the work of Cisternino et al. \cite{Cisternino} and allowed for total encapsulation of the target powder and foil substrates.

Next, a vacuum test chamber was constructed starting with a 6-inch ConFlat 6-way cross with the top and bottom ports for pumping and venting, one side port for vacuum gauging, and the opposite side port for internally mounting the HIVIPP apparatus and HV feedthrough. The remaining two flanges, front and back, were fitted with glass viewing ports (see Fig. 2). The system is brought to high vacuum using an Edwards Turbo-Pump station capable of automatic operation. During deposition, the chamber pressure is monitored using an Edwards Ion Gauge, which is usually operated in the 1.33 \textmu Pa vacuum range. The high voltage was supplied via a high voltage feedthrough to the vacuum chamber using a Spellman 20kV HV Power Supply.      

To induce deposition, the typical range of voltage starts from about 5 kV and increases to 20 kV. However, achieving these highest voltages means a greater risk of an electrical breakdown, resulting in arcing. A high vacuum environment in the chamber and employing a slow ramp to the desired voltage helps to prevent this from occurring. Deposition times can range from several hours to days. After completion, the deposited target foil backings can be removed immediately upon venting of the vacuum chamber to dry nitrogen.

\subsection{Sample Preparation}
Most of the samples encountered, and specifically the isotope, are usually supplied in powder form (+100 mesh). However, to investigate the HIVIPP deposition process, it was desired to maintain a constant and uniform size to these starting materials. The literature [3] noted the importance of particle size for the powder acceleration and movement. A simple ball milling technique was employed using a Spex GenoLyte Model 1200 shaker and for each attempt, individual milling cells for that specific target powder were used to prevent cross-contamination. The milling medium was a 7.9 mm diameter tungsten carbide ball, resulting in a fine powder for each target as starting material. Studies to characterize the particle size distribution versus quality of the deposit and minimum voltage for particle motion should be carried out in the future. 
\newline Initially, for each plating deposition, a weighed amount of sample was placed in the cylinder setup in order to gauge usage per target run. Subsequently, an excess of starting material was employed to achieve the thickest targets. For natural elements, there was always an abundance of powder samples.

\subsection{Target Substrates}
Little preparation was necessary for the foil electrodes, acting as target substrates, as these foils are already available in the target laboratory. In some instances and for actual specified targets, these substrate foils were rolled to thickness via the mechanical pack rolling technique \cite{Karasek}. After rolling and before installation into the apparatus, the foil surfaces were gently cleaned using acetone and/or ethanol. Target thickness was determined by gravimetric analysis, which is the weighing of the substrate foils before and after plating.

\section{Results}
Following the literature references, initial testing using this HIVIPP apparatus was accomplished for Carbon \cite{Sugai12C} and Titanium \cite{Skliarova} target deposits with only moderate success due to vacuum issues. For this reason, target thickness determinations were not always carried out, and some of the depositions were carried out over rather lengthy periods, as noted. Part of the problems encountered were centered on the maximum voltages attempted before breakdown being too low to induce sufficient plating. Once the vacuum-related issues were resolved, and a switch was made to an adequate high voltage supply (Glassman Series ER voltage/current limited power supply rated to 30 kV), investigations moved next to other elements, isotopically enriched metals, and proposed hazardous/radioactive material. Table I gives these results specifically for carbon target preparation, and Table 2 shows some preliminary findings for other targets that have been attempted so far. It should be noted that during these studies,
\begin{enumerate}
    \item the substrate foils are sometimes not evenly deposited
    \item top or bottom foils are sometimes not plated at all
    \item the deposition can differ depending on whether the top or bottom foil is held at ground potential.
\end{enumerate}
Further investigations to follow up on these observations are currently underway. 
\subsection{Targets Produced}
\begin{enumerate}
\item Carbon: Initial studies were undertaken on a copper substrate with a simple carbon target. The activated carbon powder chosen was already quite fine and not milled further. The copper foil substrates were cleaned with nitric acid and rinsed with distilled water. With a starting voltage of 5 kV, the particle vibration was already readily apparent, as observed with a simple laser pointer. The deposition process was left running for a week, and the targets were removed and observed to display only thin deposits on both foils. Gravimetric thickness determinations were not undertaken for these targets. Successive attempts at 8 kV were completed next on copper and also gold foil substrates, yielding usable targets. Next, carbon plating attempts were made onto thin substrates of copper, nickel, and tantalum to ascertain target thickness via energy loss measurements (see Table 1). It is unclear why these attempts made on very thin foil backings resulted in the destruction of these foils (see Fig. 3). These thin foils were probably eroded due to the vibrating bombarding particles and perhaps even become incorporated into any subsequent target deposit. This would be consistent with the observation of contamination by backing material even throughout the deposit after etching off the backing, as described by Sugai in Ref. \cite{sugai_new}.

\item Carbon-13: Isotopic carbon was next investigated for a set of requested C-13 target deposits needed for an experiment involving neutron-induced cross-section measurements. Some difficulties were encountered with the isotopic carbon powder, which prompted a pre-baking procedure in a vacuum evaporator using a platinum crucible held at 773 Kelvin for two hours. The substrates employed were specially fabricated copper disks \cite{comm}. These were relatively large and required some effort to be included in the HIVIPP setup. The mass of these disks precluded any attempt at obtaining target thickness by gravimetric methods (see Table 1). The very fine powder tended to adhere to the quartz cylinder after some time opening a path for current between the electrodes.
\item Boron: There were difficulties encountered using Boron as it is a poor conductor at standard temperatures \cite{boron} even after milling. This could also be due to the crystalline starting powder, especially with the isotopic samples. Substrates of aluminum, copper and gold were each tried. In the end, one usable B-11 target was obtained—this work continues in a forthcoming publication.
\item Titanium: Depositions with titanium were extensively investigated once the completed system was operational. The Ti powder quickly begins vibrating even at voltages as low as 5 kV and durations of 2 hours. Substrates of aluminum, copper and gold yielded good targets (see Table 2).
\item Arsenic: Arsenic can be a difficult target preparation, so HIVIPP potentially becomes an attractive method. Due to its hazardous nature, the ball milling and enclosed target cell loading were carried out in a vented chemical fume hood. Plating was accomplished on both aluminum and copper metal backing foils, resulting in usable targets for experiments as Arsenic is mono-isotopic (see Fig. 4). 
\item Zirconium: As a refractory metal, Zr was chosen as a test case. Thin deposits were easily prepared on gold substrates.

\item Tellurium	Isotopic tellurium targets on Al substrates produced by this method could be immediately deployed for medical isotope development. Natural tellurium milled powder was used as a starting material. Problems occurred with the tellurium depositing on the quartz glass cylinder, leading quickly to HV breakdown (see also Carbon-13). Nevertheless, it was shown that thick targets could be achieved.
\item Tantalum:	Another refractory metal, thick Ta deposits were prepared on Cu backings.
\end{enumerate}
\subsection{Future Prospects}
HIVIPP has proven useful for consistently making targets thicker than 100 \textmu g/cm \textsuperscript{2} and easily into the mg/cm \textsuperscript{2} range. A newer design is looking to incorporate the addition of adjustable electrodes in the chamber, allowing pumping or venting quickly inside the glass cylinder. Also desirable are electrode connections to both top and bottom substrates and computer controlled voltage. 
\newline Using HIVIPP to produce targets of the refractory metals, Molybdenum, Tungsten, Rhenium, and others could very well become the method of choice for backed targets. Investigations into the preparation of self-supporting foils by etching the substrate after plating are also being pursued. Isotopic carbon targets have already been successful, and Carbon-14 target depositions should be tested in the near future. Beryllium is another example of a hazardous substance to be tested.
\section{Conclusions}
In conclusion, the initial application of the HIVIPP method for making accelerator targets has been quite successful. Many usable targets have been prepared on several metal backings, exhibiting the versatility of the HIVIPP method. It has become a promising new technique CATS will employ for producing a variety of difficult targets, both using natural and isotopic starting powders and future applications using radioactive materials.

\begin{table}
    \centering
    \begin{tabular}{|p{0.15\linewidth} | p{0.15\linewidth} | p{0.16\linewidth} | p{0.15\linewidth} | p{0.15\linewidth}|} \hline
     
      \textbf{POWDER WEIGHT} &  \textbf{SUBSTRATE} &  \textbf{TOP/BOTTOM} &  \textbf{DURATION} &  \textbf{THICKNESS} \\ \hline
     Natural Carbon &  &  &  &  \\  \hline
      & Cu &  & 5 kV for 5 days &  \\  \hline
      &  &  & 8 kV for 7 days &  \\  \hline
      &  & Top & 8 kV for 5 days & 555 \textmu g/cm$^2$ \\  \hline
      &  & Bottom &  & 1.71 mg/cm$^2$ \\  \hline
      & Au & Top & 8 kV for 7 days & 255 \textmu g/cm$^2$ \\  \hline
      &  & Bottom &  & 628 ug/cm$^2$ \\  \hline
      3 mg & Ta & Top & 10 kV for 1 Hr & 475 \textmu g/cm$^2$ \\  \hline
      &  & Bottom &  & 350 ug/cm$^2$ \\  \hline
      8.5 mg & Ta & Top & 15 kV for 3 Hrs &  \\  \hline
      &  & Bottom &  & 5.16 mg/cm$^2$ \\  \hline
      5.3 mg & Ta & Top & 12.5 kV for 1.5 Hrs &  \\  \hline
      & Thin Cu & Bottom &  & Destroyed \\  \hline
      5 mg & Ta & Top & 15 kV for 4 Hrs &  \\  \hline
      & Thin Ni & Bottom &  & Destroyed \\  \hline
      &  &  &  &  \\  \hline
      Carbon-13 &  &  &  &  \\  \hline
      4 mg & Ta & Top & 7.5 kV overnight &  \\  \hline
      & Cu & Bottom &  &  \\  \hline
      7.2 mg & NIS Cu &  & 7.5 kV for 3 Hrs &  \\  \hline
      & Ta & Bottom &  &  \\  \hline
      Baked C-13 & NIS Cu & Top & 7.5 kV for 4 Hrs &  \\  \hline
      & Ta & Bottom &  &  \\  \hline
    \end{tabular}
    \caption{Carbon Targets Prepared Using HIVIPP}
    \label{tab:AlphaCounts}
\end{table}

\begin{table}
    \centering
    \begin{tabular}{|p{0.15\linewidth} | p{0.15\linewidth} | p{0.16\linewidth} | p{0.15\linewidth} | p{0.15\linewidth}|} \hline
     
      \textbf{POWDER WEIGHT} &  \textbf{SUBSTRATE} &  \textbf{TOP/BOTTOM} &  \textbf{DURATION} &  \textbf{THICKNESS} \\ \hline
      Boron & Al,Cu &  &  & failed \\  \hline
      B-10 & Au & Bottom &  & 102 \textmu g/cm$^2$ \\  \hline
      Titanium & Al &  & 5 kV for 2 Hr &  \\  \hline
      & Cu &  &  &  \\  \hline
      & Au & Top &  & 820 \textmu g/cm$^2$ \\  \hline
      &  & Bottom &  & 1.07 mg/cm$^2$ \\  \hline
      Arsenic & Al & Top &  & 1.93 mg/cm$^2$ \\  \hline
      &  & Bottom &  & 5.13 mg/cm$^2$ \\  \hline
      & Cu & Top &  & 1.93 mg/cm$^2$ \\  \hline
      &  & Bottom &  & 1.53 mg/cm$^2$ \\  \hline
      Zirconium & Au & Top &  & 254 \textmu g/cm$^2$ \\  \hline
      &  & Bottom &  & 351 \textmu g/cm$^2$ \\  \hline
      Tellurium & Al & Top &  & 2.57 mg/cm$^2$ \\  \hline
      &  & Bottom &  & 45 \textmu g/cm$^2$ \\  \hline
      &  & Bottom &  &2.72 mg/cm$^2$ \\  \hline
      Tantalum & Cu & Top &  & 9.79 mg/cm$^2$ \\  \hline
      &  & Bottom &  & 10.15 mg/cm$^2$ \\  \hline
    \end{tabular}
    \caption{Non-Carbon Targets Prepared Using HIVIPP}
    \label{tab:AlphaCounts}
\end{table}

\begin{figure}[h]
    \centering
    \includegraphics[scale=1]{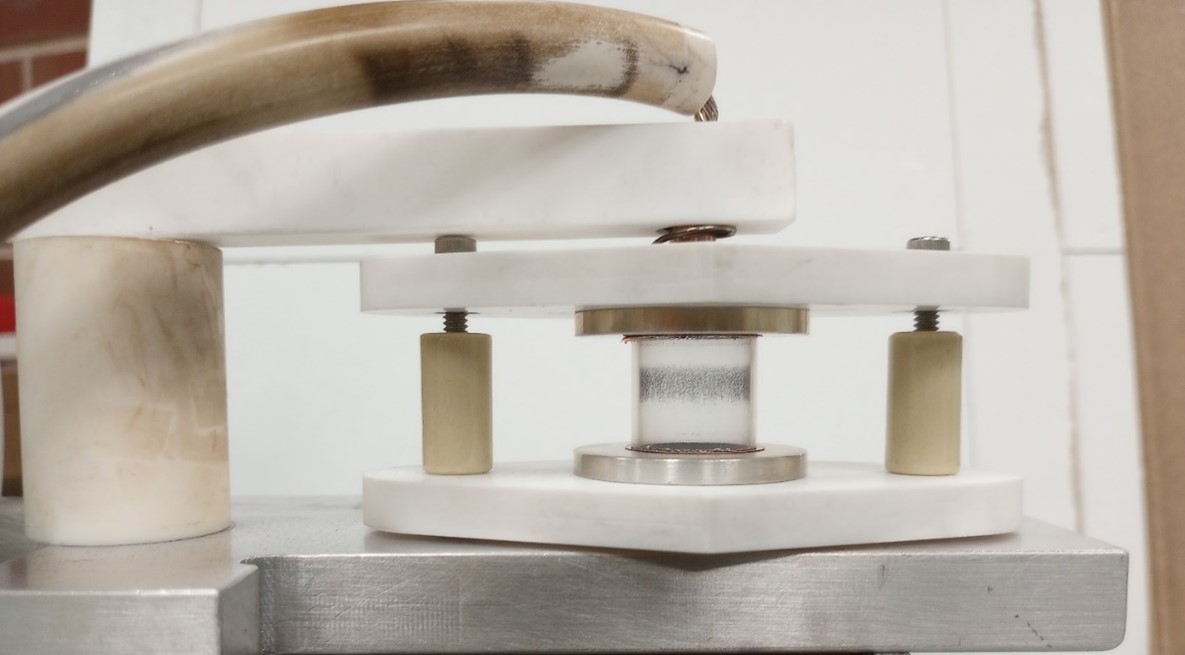}
    \caption{Photograph showing a close-up of the HIVIPP cell encapsulation design}
    \label{fig:HIVIPPCloseUp}
    
\end{figure}

\begin{figure}[h]
    \centering
    \includegraphics[scale=1]{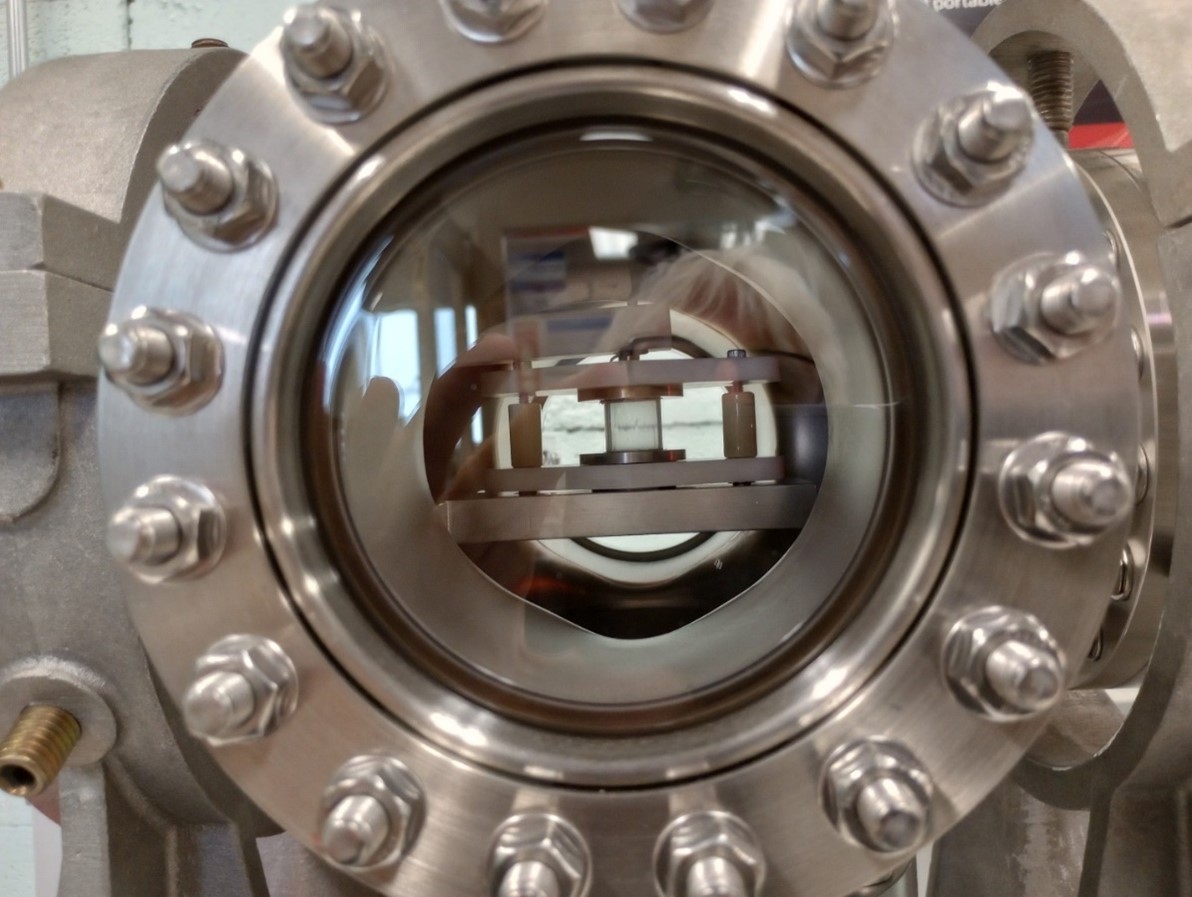}
    \caption{Photograph showing the plating cell arrangement inside the HIVIPP Apparatus}
    \label{fig:HIVIPP8-wayCross}
    
\end{figure}

\begin{figure}[h]
    \centering
    \includegraphics[scale=1]{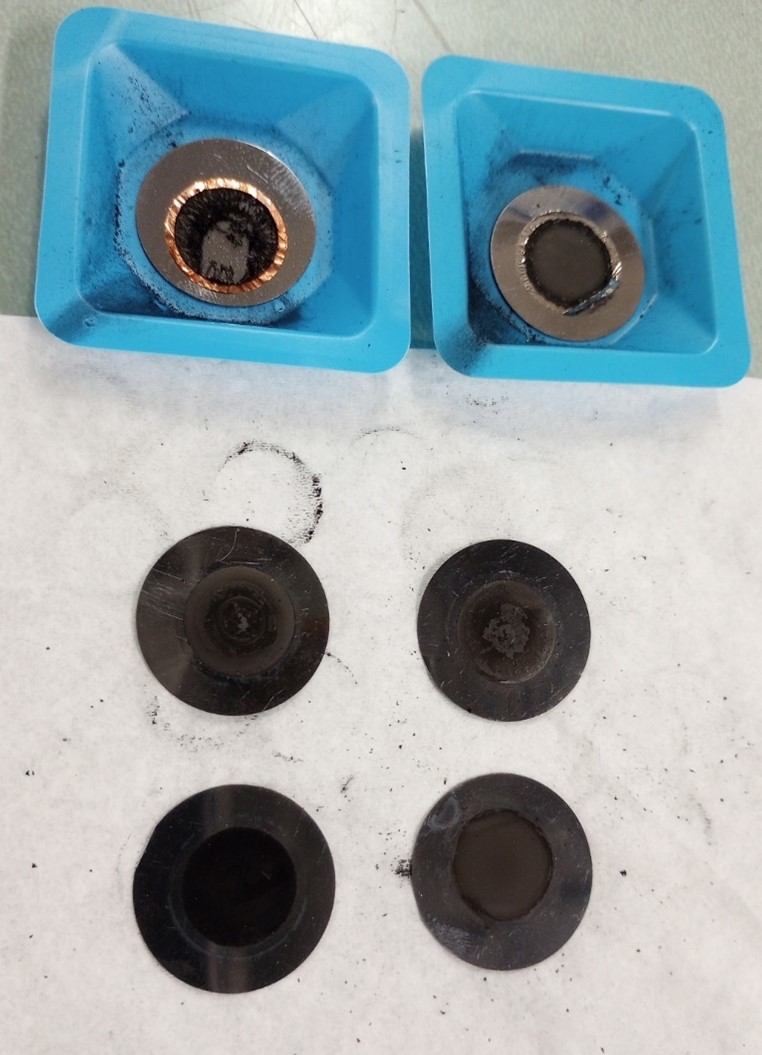}
    \caption{Several attempts at carbon deposition onto thick Ta backing as well as thin Ni and Cu substrates (shown as completely disintegrated?)}
    \label{fig:CarbonDeposition}
    
\end{figure}

\begin{figure}[h]
    \centering
    \includegraphics[scale=1]{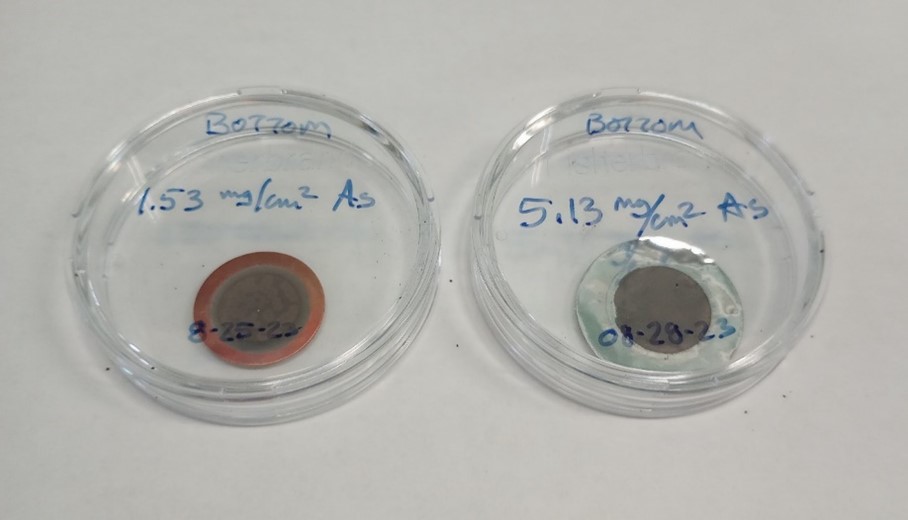}
    \caption{Arsenic Targets deposited onto Al and Cu backings}
    \label{fig:ArsenicDeposition}
    
\end{figure}


\acknowledgments
This material is based upon work supported by the U.S. Department of Energy, Office of Science, Office of Nuclear Physics, under Contract No. DE-AC02-06CH11357. This research used resources of ANL’s ATLAS facility, which is a DOE Office of Science User Facility


{\footnotesize                                              
    \bibliographystyle{ieeetr}                              
    \bibliography{main}
    }
\end{document}